\documentclass[%
 reprint,
 amsmath,amssymb,
 aps,
]{revtex4-1}

\usepackage{graphicx}
\usepackage{dcolumn}
\usepackage{bm}

\usepackage[utf8]{inputenc}
\usepackage[T1]{fontenc}
\usepackage{graphicx}
\usepackage{braket}
\usepackage{dsfont}
\usepackage{amsmath} 
\usepackage{natbib} 
\usepackage{textcomp}
\usepackage[utf8]{inputenc}

\begin{document}


\title{The Minimum Distance of PPT Bound Entangled States from the Maximally Mixed State.}

\author{Shreya Banerjee
}
 \email{shreya93ban@gmail.com}
 \affiliation{Indian Institute of Science Education and Research Kolkata, \\
Mohanpur- 741246, West Bengal, India\\
}


\author{Aryaman A. Patel}%
 \email{13me121.aaryaman@nitk.edu.in}
\affiliation{National Institute of Technology Karnataka,  Surathkal\\ Mangalore-575025, Karnataka, India\\}

\author{Prasanta K. Panigrahi}%
 \email{panigrahi.iiser@gmail.com}
\affiliation{Indian Institute of Science Education and Research Kolkata, \\
Mohanpur- 741246, West Bengal, India\\}

\begin{abstract}
Using a geometric measure of entanglement quantification based on Euclidean distance of the Hermitian matrices \cite{patel2016geometric}, we obtain the minimum distance between a bipartite bound entangled $n$- qudit density matrix and the maximally mixed state. 
This minimum distance for which entangled density matrices necessarily have positive partial transpose (PPT) is obtained as $\frac{1}{\sqrt{\sqrt{d^n(d^n-1)}+1}}$, which is also a lower limit for the existence of 1-distillable entangled states. The separable states necessarily lie within a minimum distance of $\frac{R}{1+d^{n-1}}$ from the Identity \cite{patel2016geometric},where R is the radius of the closed  ball homeomorphic to the set of density matrices, which is lesser than the limit for the limit for PPT bound entangled states. Furthermore an alternate proof on the non-emptiness of the PPT bound entangled states has also been given. 

\end{abstract}

\pacs{Valid PACS appear here}
\maketitle


\section{\label{sec:level1}INTRODUCTION}
Characterization of entanglement is of deep interest in the field of quantum information and quantum computation \cite{bruss2002characterizing,PhysRevLett.103.240502,jaeger2009entanglement}. As is well known there are two types of entangled states: distillable and non-distillable \cite{PhysRevA.53.2046}. Distillable  entangled states find application in quantum technology: quantum teleportation \cite{PhysRevLett.76.722,Chen2017}
,quantum error correction \cite{imai2006special,Lipka-Bartosik2016}, quantum cryptography \cite{PhysRevLett.94.040503,acin2003security,Gao2017} etc. The other class of entangled states that cannot be distilled is called bound entangled states, which has found application in steering and ruling out local hidden state models \cite{PhysRevLett.113.050404}. The Peres-Horodecki criterion provides a necessary and sufficient condition for separability in $2\otimes 2$ and $2\otimes 3$ dimensions. It fails to identify separable states in higher dimensions. Using this criterion, one can only identify the states that have positive partial transpose (PPT) in higher dimensions. By definition, such states are definitely bound entangled states \cite{PhysRevA.61.062312}. There was no straightforward method to separate bound entangled states as a class. A deeper understanding of these class of states is thus of high importance both from fundamental and application perspectives. 

 There have been various approaches to analyze the geometry of the quantum state space \cite{braunstein1995geometry,zyczkowski2001induced,zyczkowski2001monge} and entanglement measures based on geometry \cite{patel2016geometric,ozawa2000entanglement,heydari2004entanglement,PhysRevLett.77.1413,goswami2017uncertainty}. Geometry has also been used in quantum computation to form new algorithms on \cite{LaGuardia2017,Holik2017}. Recently quantification of entanglement has been carried out from a geometric perspective, for general n qudit states \cite{patel2016geometric}. There is also another approach using wedge product which manifests naturally in a geometric setting \cite{bhaskara2017generalized}. This geometric approach makes essential use of the fact that measurement of a subsystem of an entangled state necessarily affects the remaining constituents in contrast to separable states. Using the geometry of $N=d^n$- dimensional positive semidefinite matrices, here we establish a criterion for arbitrary dimensions for separating  PPT bound entangled states. Interestingly this class of states can be associated to almost every pure entangled states \cite{PhysRevA.75.012305}. Previously the lower limit for separable states has been established in ref. \cite{patel2016geometric}. As is well known the PPT criterion is useful for  dimensions less than equal to $6$. Here we provide the geometric lower bound for arbitrary dimensions within which every state is PPT. For dimensions greater than $6$ it gives the  minimum distance from the maximally mixed state for which a state is bound entangled.  
 
 The paper is organized as follows: Sec. II describes classification of states based on their partial transpose and the general geometry of density matrices. The spectrum of the partially transposed matrix of a pure state is discussed in Sec. III. In Sec. IV the boundary of the PPT bound entangled states have been calculated along with an alternate proof of the non-emptiness of the PPT bound entangled states. We then conclude in Sec. V with directions for future work.

\section{CLASSIFICATION AND MEASUREMENT BASED GEOMETRY OF N DIMENSIONAL DENSITY MATRICES }
A general state $\rho$ acting on $H_A \otimes H_B$ can be written as \cite{djokovic2016two},
\begin{equation}
    \rho= \sum_{ijkl}p^{ij}_{kl} \ket{i}\bra{j}\otimes \ket{k}\bra{l},
\end{equation}
with its partial transpose defined as,
\begin{equation}
    \rho^{T_B}=\mathds{I}\otimes T(\rho)=\sum_{ijkl}p^{ij}_{kl} \ket{i}\bra{j}\otimes \ket{l}\bra{k}.
\end{equation}
Here $\mathds{I}\otimes T(\rho)$ is the map that acts on the composite system with Identity map acting on system A and transposition map acting on B.

 $\rho $ is called PPT if its partial transpose $\rho^{T_B}$ is a positive semi-definite operator. If  $\rho^{T_B}$ has a negative eigenvalue, it is called NPT. It is known from Peres-Horodecki criterion that for $2\otimes2$ and $2\otimes3$
dimensions, all PPT states are separable and all NPT states are entangled. For arbitrary $n\otimes m$ dimensions, some PPT states show entanglement,whereas all NPT states are necessarily entangled \cite{djokovic2016two}. 

For a bipartite state $\rho$ , acting on $H=H_A\otimes H_B$ and for an integer $k\geq1$, $\rho $ is k-distillable if there exists a (non-normalised) state $\ket{\psi}\in H^{\otimes k} $ of Schmidt-rank at most 2 such that,
 $$ \bra{\psi}\sigma^{\otimes k} \ket{\psi}<0 , \sigma=\mathds{I}\otimes T(\rho).$$
 $\rho$ is distillable if it is k-distillable for some integer $k \geq 1$ \cite{PhysRevA.61.062312}.

 If a state $\rho$ is PPT, it is non-distillable, hence entangled PPT states have no distillable entanglement. Such states are called PPT bound entangled states. All distillable entangled states are NPT. The converse may not hold, i.e. if all NPT states are distillable or not. It is believed that the converse does not hold \cite{PhysRevA.61.062312} .
 
 \begin{figure}[h!]
\centering
\includegraphics[scale=0.44]{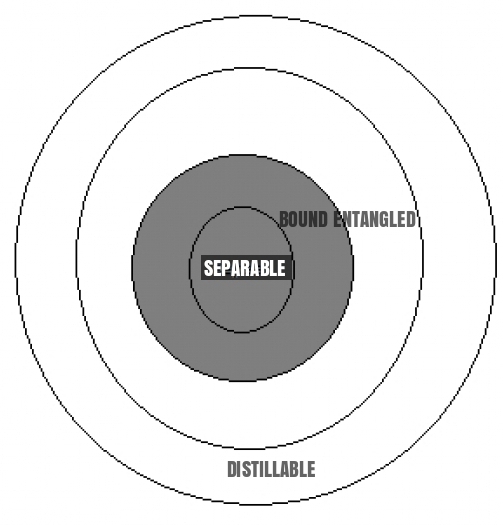}
\caption{Diagrammatic representation of the set of all mixed states, for a general
arbitrary Hilbert space; the shaded region is the set of all PPT states and the white portion is the set of all NPT states.}
\end{figure}

The Euclidean distance between any two  Hermitian matrices $\rho$ and $\sigma$ is given by \cite{patel2016geometric},

\begin{equation}
    D(\rho , \sigma)= \sqrt{Tr{(\rho-\sigma)}^2}.
\end{equation}

The set of all density matrices of order N is considered as a convex compact set embedded in the closed ($N^2-1$) ball $\mathds{B}^{N^2-1}$ of radius $\sqrt{\frac{N-1}{N}}$, centred at normalised identity $\frac{\mathds{I}}{N}$. This set always admits a regular N-1 simplex as one of its orthogonal basis. The convex hull of a basis is represented by a regular n simplex centred at normalised identity $\frac{\mathds{I}}{N}$ and circumscribed by $\mathds{B}^{N^2-1}$, where 
$N-1\leq n \leq N^2-1$. Each density matrix can be treated as a point in a simplex whose vertices are pure states.

\begin{figure}[h!]
\centering
\includegraphics[scale=0.85]{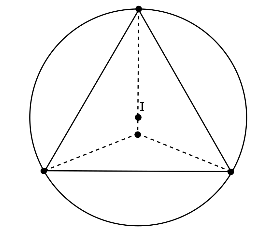}
\caption{Orthogonal basis represented by a 3-simplex, a regular
tetrahedron, for N=4 case.}
\end{figure} 

Using this geometry, it has been shown that the set of all n qudit density matrices, whose distance from $\frac{\mathds{I}}{N}$ is less or equal to $\frac{1}{1+d^{n-1}}\sqrt{\frac{d^n-1}{d^n}}$, are separable \cite{patel2016geometric}.

A bipartite $n$ qudit density matrix $\rho$ with bi-partitions A-B is considered.
To find out the separability criterion for  $\rho$, a measurement on one of the bipartitions is done \cite{patel2016geometric}. Then one checks if both the post-measurement reduced density matrices $\rho_A$ and $\rho_B$ localizes to the simplices of corresponding dimensions. The distance between the reduced density matrices and the centre of the closed ball homeomorphic to the corresponding simplex is measured using Eq. 3 and if it lies within the bound given in ref. \cite{patel2016geometric} then it is certainly separable. A similar approach has been used in \cite{bhaskara2017generalized} where a bipartite $n$ qudit pure state is projected in a basis consisting two orthonormal bases to check the separability of the state.

 A subset S of a vector space V is called a cone if $\forall x \in C$ and positive scalar $\alpha$, $\alpha x \in C$. A cone C is called a convex cone if $\alpha x + \beta y \in C $. 

The defining property of the set of all  $N \times N$ positive semidefinite matrices P is that the scalar $x^T x$ is positive for each nonzero coloumn vector x of N real numbers. If P be the set of all symmetric positive semi-definite matrices, then  $\forall X, Y \in P $ and $ \alpha,\beta > 0 $,\\ $ x^T(\alpha X + \beta Y)x=\alpha x^T Xx+ \beta x^T Yx > 0 $ ,i.e., P is a convex cone.

The set of the symmetric positive semidefinite(PSD) matrices of order $N\times N$ forms a convex cone $S_N$ in $\mathds{R}^{N^2}$. A few interesting properties of this cone are, 

(a) it has non-empty interior containing positive definite matrices which are full rank,

(b) on the boundary of the cone the singular positive semidefinite matrices with at least one eigen-value zero lie. 

The origin of this cone is identified as the only matrix with all eigenvalues zero which is equidistant from each point on the surface of the $\mathds{B}^{N^2-1}$ ball.This is only possible when the ball is embedded in a subspace of the cone. The intersecting region of the ball and the cone is then in a dimension $N-1$. 

Taking a transposition of one of the subsystems if the post-transposition density matrix lies within the cone formed by the positive semidefinite matrices, then they are assumed  to be PPT. 

Each $N\otimes N$ positive semidefinite matrix is associated with a quadric. One can represent a diagonalised symmetric matrix of order $2\otimes 2$ as a conic using 
the characteristic equation of the matrix. Let us consider the matrix 
\[
\begin{bmatrix}
a & 0\\
0 & b

\end{bmatrix}
\]

where a is an eigenvalue of the matrix with respect to the eigenvector
\[
\begin{bmatrix}
x_{1}\\
x_{2}
\end{bmatrix}
\]
The corresponding equation of conic will be,
\begin{equation}
    a{x_1}^{2}+b{x_2}^{2}=a
\end{equation}

Each positive definite diagonalised matrix in $3\otimes 3$ dimension will form an ellipsoid and each positive semidefinite matrix will either be a set of intersecting planes or parallel planes. The origin with all three eigenvalues zero will give a point.  


The set of $n$ qudit density matrices is represented by convex sets homeomorphic to a closed ball of radius $R$ centred at the maximally mixed state, the identity matrix of $d^n\otimes d^n$ dimension. Expectedly this contains entangled states with positive and negative partial transpose. If the partial transpose of a density matrix is positive, then it will lie within the $S_N$ cone. Now the minimum distance of the $S_N$ cone from the maximally mixed state placed at the centre of the $\mathds{B}^{{N^2}-1}$ ball is the distance for which the density matrix would definitely be PPT.   

\section{SPECTRUM OF THE PARTIALLY TRANSPOSED MATRIX OF A PURE STATE}

The spectrum of the partial transposition of a pure state has been given in \cite{PhysRevA.58.883}.
We consider the density matrix of a pure state  $\rho=\ket{\psi}\bra{\psi}$.

The Schmidt decomposition of  $\ket{\psi} \in H= H^m \otimes H ^n$
 is given by,
 \begin{equation}
     \ket{\psi}=\sum_i \alpha_i \ket{e_i}\otimes\ket{f_i}
 \end{equation}
where $\ket{e_i} \otimes \ket{f_i}$ forms a bi-orthogonal basis, i.e., $\bra{e_i}\ket{e_j}=\bra{f_i}\ket{f_j}=\delta_{ij}$ and $0\leq \alpha_i \leq 1$ along with $\sum_i \alpha _i^2=1$.
\\[5 mm]
The partial transposition of $\rho$, $\rho^{T_B}$ has eigenvalues, $\alpha_i^2$ for i=1,2,....r where r is the Schmidt rank; $\pm\alpha_i\alpha_j$ for $1\leq i < j \leq r$ and 0 with multiplicity $min(m,n)\lvert m - n\rvert +{ \{ min (n,m) \} }^2 -
r^2$.

Also, all eigenvalues of partial transposition of any $m\otimes n$ state always lie within [-1/2, 1] \cite{PhysRevA.87.054301}.
\section{}
\subsection{DISTANCE OF THE
PARTIAL-TRANSPOSE OF A N
DIMENSIONAL MATRIX FROM
NORMALISED IDENTITY}

We consider a bipartite $n$ qudit density matrix $\rho$. If the partial transpose of $\rho$, $\rho^{T_B}$ is positive semidefinite, then $\rho$ is PPT. One can infer that the minimum distance between $S_N$ cone and the centre of the $\mathds{B}^{N^{2}-1}$ ball is the lower limit of the distance between $\rho^{T_B}$ and the maximally mixed state.

Any density matrix $\rho$ of order $N \times N$ can be written as \cite{PhysRevLett.80.2261},

\begin{equation}
\rho = pP_{\psi}+(1-p)\rho^{//},   
\end{equation}

where $p \in[0, 1]$; $P_\psi$ is a pure state and $\rho^{//}$ any density matrix.

Partial transpose followed by diagonalisation of Eq. 6, yields,

\begin{equation}
    \sigma^{T_B}=p(\sigma^{T_B})^/+(1-p)({\sigma^{T_B}})^{//}, 
\end{equation}

where  $\sigma^{T_B}$ is the diagonalised partial transpose of $\rho$, $(\sigma^{T_B})^/$ is the
diagonalised partial transpose of $P_{\psi}$ and $({\sigma^{T_B}})^{//}$ is the diagonalised partial transpose of $\rho^{//}$.

As  $\sigma^T_B$  is diagonal, it is also symmetric. Therefore if  $\sigma^T_B$ is positive
semi-definite or positive definite, it lies either on the boundary of the convex
cone $S_N$ or inside it.

We consider the Euclidean distance between  $\sigma^{T_B}$ and normalised identity of order N: 

\begin{equation}
    D( \sigma^{T_B}, \mathds{I}_N)=\sqrt{Tr{( \sigma^{T_B}-\frac{\mathds{I}}{N})}^2}
\end{equation}

One obtains from Eq. 8,

\begin{equation}
    D( \sigma^{T_B}, \mathds{I}_N)=\sqrt{Tr{(\sigma^{T_B}})^2-\frac{2}{N}Tr(\sigma^{T_B})+\frac{1}{N}}
\end{equation}

Substituting the value of  $\sigma^{T_B}$ from Eq. 7 and using the spectrum of the partially transposed matrix of a pure state,

\begin{equation}
\begin{split}
{D( \sigma^{T_B}, \mathds{I}_N)}^2=
p_{m}^2+(1-p_{m})^2\sum_j{\lambda_j}^2  
+2p_{m}(1-p_{m})\lambda_j\\-\frac{2}{N}(p_{m}+(1-p_{m})\sum_j\lambda_i)+\frac{1}{N}
\end{split}
\end{equation}

where $\lambda_j$ is the jth eigenvalue of the matrix $({\sigma^{T_B}})^{//}$, i corresponds to the only surviving eigenvalue of the partial transposed matrix of the pure state, $({\sigma^{T_B}})^{/}$ and $p_{m}$ is the maximum value of the parameter p.

Werner states of $N\otimes N$ dimensions can be written as \cite{patel2016geometric},
\begin{equation}
    \rho_{w}= pP_{\psi}+(1-p)\frac{\mathds{I}}{N},
\end{equation}

where $p \in[0, 1]$; $P_\psi$ is a pure state and $\frac{\mathds{I}}{N}$ is the normalised Identity matrix of order N.

The normalised identity of order N is the partial transpose of itself.
Substituting the values of $\sum \lambda_j$ and $\lambda$ in Eq. 10 considering $\lambda_j$ as the eigen value of $\mathds{I}_N$ one can have the distance of the partially transposed Werner states from the maximally mixed states as,

\begin{equation}
    D= p_{m}\sqrt{\frac{N-1}{N}},
\end{equation}
  where $p_{m}$ is the maximum value of the parameter p.
  
  Calculating the distance between $\rho_w$ and normalised identity using Eq. 3, we reach to Eq. 12 for the distance of partially transposed Werner states from the maximally mixed states.

\subsection{MINIMUM DISTANCE FOR WHICH A STATE WOULD BE PPT BOUND ENTANGLED}

We consider a density matrix of order $4$ and bi-partitions $2\otimes 2$. In this case there is no bound-entanglement as the PPT criterion is necessary and sufficient for separability for $2\otimes 2$ systems. If the density matrix is PPT then the partial transpose of the matrix will lie within the cone $S^4$ of all positive semidefinite matrices of order $4$. The set of all density matrices is homeomorphic to the closed ball $B^{15}$ and the cone $S^4$ intersects it in a $3$-dimensional space.

The boundary of $S^4$ is formed by positive semidefinite matrices of order $4$, which are a set of parallel planes and in their $2$ dimensional projections they form a set of parallel lines ($x^2=a^2$, where a is a non-zero eigenvalue of the system) or a set of intersecting lines ($\frac{X^2}{a^2}=\frac{y^{2}}{b^{2}}$, where a and b are both non zero eigenvalues of the system). The parallel lines intersect the $3$ dimensional sphere of the ball at the surface, at  [1 0 0], [0 1 0] and [0 0 1], namely the pure states. The intersecting lines give an idea of the positive partially transposed mixed states.

The equation of the intersecting lines is,
\begin{equation}
    \frac{x^2}{a^2}=\frac{y^2}{b^2}
\end{equation}

 where a and b are eigenvalues of the corresponding system.

Equation $13$ reduces to,

\begin{equation}
   x=\pm{\frac{a}{b}}y
\end{equation}

Fig. 3 depicts that,


\begin{figure}[h!]
\centering
\includegraphics[scale=0.45]{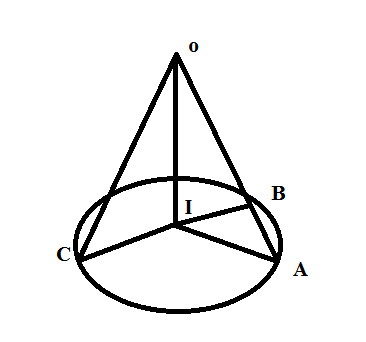}
\caption{Cross-section of the $\mathds{B}^{15}$ ball and $S_4$ cone, showing the centre of the ball I and the origin of the cone O. }
\end{figure} 

 A and C are the points where the cone cuts the ball and IB is the minimum distance from I to the boundary of ball. The slope of the intersecting lines OC and OA is giveb by $\frac{a}{b}$. Slope of OA is also given by $\frac{OI}{IB}$ as O is perpendicular to the plane where I lies.

The distance of the maximally mixed state from the origin of the $S^4$ cone $OI=\frac{1}{\sqrt{4}}$.

$IB$ is the minimum distance of the cone from the maximally mixed state. 

The slope of the line OA is given by,
\begin{equation}
    \frac{b}{a}=\frac{OI}{IB},
\end{equation}

\begin{equation}
    IB=\frac{b}{a\sqrt{4}}.
\end{equation}

The minimum value of $\frac{b}{a}$ is,
\begin{equation}
    {\frac{b}{a}}_{min}=\frac{\frac{1}{\sqrt{\lambda_{max}}}}{\frac{1}{\sqrt{\lambda_{min}}}}=\frac{\sqrt{\lambda_{min}}}{\sqrt{\lambda_{max}}}
\end{equation}

and  minimum value of IB is obtained as,
\begin{equation}
    IB_{min}=\sqrt{\frac{\lambda_{min}}{\lambda_{max}}}\frac{1}{\sqrt{4}}
\end{equation}

In this case the intersecting part of the $S^4$ cone and the $B^{15}$ ball is the intersecting part of a 3-d image of the cone and the $B^3$ ball, which is a Bloch sphere. Hence each matrix in the intersection is a density matrix. Therefore,

\begin{equation}
    \sqrt{\lambda_{min}}=\sqrt{\frac{1}{4}}
\end{equation}
 and,
 
 \begin{equation}
      \sqrt{\lambda_{max}}=\sqrt{\frac{3}{4}}.
 \end{equation}

and,
 
\begin{equation}
    IB_{min}=\frac{1}{\sqrt{12}}
\end{equation}

The ratio at which this value of IB cuts a proper radius of $B^{15}$ ball corresponds to the value of $p_{m}$ for Werner sates. This value is given by,

\begin{equation}
   p_{m}=\frac{IB}{R_{15}}=\frac{1}{3}
\end{equation}

PPT criterion is necessary and sufficient for separability in a bipartite $2\otimes 2$ systems. Following that a $4$ dimensional state is absolutely separable if it lies within a distance of $\frac{1}{3}R$ of the maximally mixed state. This result matches with the separability criterion known for the 4-qubit Werner states \cite{patel2016geometric}.

For higher dimensions PPT criterion is not sufficient for separability. Instead, the criterion helps us to detect bound entangled states. For  $N$ dimensional states the cone of all PSD matrices intersects the $B^{N^2}$ ball in $N-1$ dimensions. Considering the geometry of diagonalised PSD matrices of $N-1$ dimensions one can say that they are associated with either $N^/$ dimensional ellipsoids that form the curved surface of the cone where $3\leq N^{/}\leq (N-3)$ or intersecting lines that give the boundary of the cone or intersecting planes  which denotes the points where the cone cuts the ball. 

Considering the intersecting lines, we have the minimum distance IB of the boundary of the cone from the maximally mixed state as, 

\begin{equation}
    IB=\sqrt{\frac{\lambda_{min}}{\lambda_{max}}}\frac{1}{\sqrt{N}}
\end{equation}

This minimum distance satisfies the following inequalities:

\begin{equation}
    {(\lambda_1-\frac{1}{N})}^2 +  {(\lambda_2-\frac{1}{N})}^2 \geq 0,
\end{equation}
 
 and,
 
 \begin{equation}
      {(\lambda_1-\frac{1}{N})}^2 +  {(\lambda_2-\frac{1}{N})}^2 \leq \frac{N-1}{N}.
\end{equation}

 $\lambda_1$ and $\lambda_{2}$ are two eigen values of the corresponding PSD matrix.

One can then obtain,
$\lambda_{min}=\frac{1}{\sqrt{N}}$ \\ and $\lambda_{max}=\sqrt{\frac{N-1}{N}}+\frac{1}{N}$
 
From Eq. 22 one can obtain,
\begin{equation}
    IB_{min}=\frac{1}{\sqrt{\sqrt{N(N-1)}+1}}
\end{equation}
 
 Comparing this distance with the distance found from Eq. 10, one can determine if a state is absolutely PPT bound entangled. The value of parameter p for which the Werner states of $N$ dimension will be PPT is also obtained as,
 \begin{equation}
     p_m=\sqrt\frac{N}{N-1}\frac{1}{\sqrt{\sqrt{N(N-1)}+1}}
 \end{equation}





 






This is the maximum value of parameter p for which the partially transposed matrix of any density matrix of order N lies within the convex cone formed by the positive semidefinite and positive definite matrices. The distance of any density matrix $\rho$ of order N from $\mathds{I}_N$ is given by, $p(\sqrt{\frac{N-1}{N}})$ [1].
For $\rho$ to be definitely PPT, the maximum distance of $\rho$ from $\mathds{I}_N$ is, 
$\frac{1}{\sqrt{\sqrt{N(N-1)}+1}}$
.

The bipartite $2\otimes 2$ systems have a different value of the distance due to the fact that there the cone cuts the closed ball at a dimension where the intersection part lie inside a Bloch sphere. All the points inside the sphere represent a quantum state, which is not true for the higher dimensional case.

For n qudit density matrices, all matrices within the distance $\frac{1}{\sqrt{\sqrt{d^n(d^n-1)}+1}}$.
from $\mathds{I}_N$ are PPT. 

It is shown in ref. \cite{patel2016geometric} that all n-qudit density matrices within distance $\frac{1}{1+d^{n-1}}\sqrt{\frac{d^n-1}{d^n}}$  from the normalised identity are separable. This implicates, all entangled n-qudit density matrices within distance $\frac{1}{1+d^{n-1}}\sqrt{\frac{d^n-1}{d^n}}$  to $\frac{1}{\sqrt{\sqrt{d^n(d^n-1)}+1}}$  are PPT bound entangled. Using the definition of 1-distillable entangled states [21], one can infer that no 1-distillable entangled states lie within this distance. The states within this distance are necessarily PPT bound entangled. This proves the non-emptiness of the set of such states.









\section{CONCLUSION}
In summary, we have used a measurement based geometric approach to check if the partial transpose of a $n$ qudit density matrix gives all non negative eigenvalues. It has been shown that all the density matrices within the distance $\frac{1}{\sqrt{\sqrt{d^n(d^n-1)}+1}}$ from the maximally mixed state have a positive partial transpose. The precise distance for which a Werner state is PPT bound entangled has also been found. 
As the lower limit for distance between maximally mixed state and a separable state \cite{patel2016geometric} is less than the distance between a PPT bound entangled state and the same found here, one can conclude that the set of PPT bound entangled states is non-empty. This limit also applies as a geometrical lower bound for 1-distillable entangled states. The method provided here may find use to calculate the limits for k-distillable states and the NPT bound entangled states.






\bibliographystyle{unsrt}

\end{document}